# Predicting the growth rate of helium bubbles in metal tritide

Fan-Xin Meng[1,2], Chen Ming[1,2], Xi Chen[1,2], Wei-Feng Yu[1,2] and Xi-Jing Ning[1,2] *

*1 Institute of Modern Physics, Fudan University, Shanghai 200433, China*

*2 Applied Ion Beam Physics Laboratory, Key Laboratory of the Ministry of Education, Fudan University, Shanghai 200433, China*

**Abstract**

Helium bubbles nucleation and growth in metals or metal tritide is a long-standing problem attracting considerable attention in nuclear industry but the mechanism remains indistinct and predicting the growth rate of helium bubble is inexistence still up to new. Here, the rate of helium bubbles nucleation and growth in metal tritide is developed based on a dynamical model, which describes the diameter of helium bubbles increasing linearly as $t^{1/3}$ in titanium tritide at room temperature, agreeing quite well with the experimental phenomenon. The way of reducing storage temperature from 300 to 225 K or increasing the helium atoms diffusion barrier from 0.81 to 1.1 eV can effectively restrain bubbles growth and prolong lifetime of titanium tritide more than 4 times, which provides a useful reference to relevant experiment exploration and applications. This model also can be used to predict lifetime of new tritium-storage materials and plasma facing materials in nuclear industry.

* Corresponding author. E-mail: xjning@fudan.edu.cn (X.J. Ning).

**Introduction**

The diffusion and accumulation of helium atoms from tritium decay form high pressure helium bubbles in metal tritide, which lead to change the properties, such as swelling, surface roughening and blistering, and eventually fracture the materials, reducing service lifetime [1-8]. Although, this phenomenon was found firstly in the mid to late 1970s [1], the microscopic mechanism of helium bubbles nucleation and growth in the matrix (what the mechanism of helium atoms diffusion is, where the bubbles nucleate and what the volume of bubbles as a function of aging time is) remains in debate and the quantitative studies of parameters which can influence it are more indistinct still up to new.

In recent several decades, in order to prolong service lifetime of metal tritide, the atomic behavior of helium and bubbles formation and growth in metal tritide has been investigated by a lot of experimental approaches and theoretical methods [9-23]. Systematic experimental studies of the transient material properties resulting from tritium decay are extremely time consuming, expensive, and difficult [9-16], because any research to map the whole life of a metal tritide needs several years and the complexity of radiological TEM sample preparation and image analysis are almost as hard as imagined. Thus the allure of theoretical investigations is to mitigate these difficulties while gaining quantitative insight into the fundamental physical processes that drive the macroscopic behavior. There are two ways to study helium bubbles formation and growth in metal tritide in theory. one is static computation at atomic level. At first, Weaver [17] believed the helium atoms are trapped at the octahedral interstitial sites in tritide with the fluorite crystal structure while Wilson [18] suggested helium atoms are able to cluster with each other, forming bubbles, in the perfect matrix by atomistic calculation, based on a sample model. Then recently, many groups have attempted to understand the atomic behavior of helium using the first principle calculation and provided the microcosmic diffusion path of helium atom [20-23]. But those studies are verified hardly by experiment, and the dynamic information about the growth of helium bubbles as function of aging time in metal tritide cannot be provided by those studies. Another is dynamic research. The molecular dynamics (MD) or Monte Carlo method seems a direct approach to the problem, but the covered timescale cannot go beyond several microseconds and provide authentic information of helium bubbles formation and growth in the metal tritide.

Actually, helium bubbles growth (or materials lifetime) prediction is not only needed for tritium storage materials, it has been a long-standing problem of plasma facing materials (metals and alloys) in nuclear reactors. In this study, our principal interests are in building the bridge that can connect microcosmic and macroscopic behavior about helium bubbles in materials, take metal tritide as, based on a statistical model, which can predict helium bubble diameter at different aging time and discuss quantitatively the parameters influenced it. Those are of crucial importance for prolonging service lifetime of materials in nuclear industry by restraining bubble formation and growth.

**Models and Methods**

In the aging process, considering the tritium decay randomly into $^3$helium and the density of isolated helium bubbles remains roughly constant in metal tritide [9, 10, 16], it was suggested that helium atoms are in uniform distribution, helium bubbles nucleate at inherent defects (such as vacancies, dislocations, grain boundaries and so on) and the density of helium bubbles is equal to that of inherent defects in metal tritide. Thus, it is regarded that the bulk metal tritide are composed of the small sphere element (gray range, radius is $r_1$) centered on inherent defects, as show in **Fig.1**, and then total generated helium atoms from tritium decay will be trapped at the defect, forming bubbles, during them diffusion in small sphere element. In this sphere element, it is worth noting that the density of helium decayed by tritium should is too low to interact on each other, forming clusters, before them were trapped at the defect during them diffusion, which can be verified by comparing with the estimated minimum density ($\rho_{Stan}$) with interaction between helium atoms. So to predict the growth rate of helium bubbles, we only need to quantitatively predict the number of trapped helium atoms ($N_{trap}$) at the defect as a function of aging time in this small sphere element of metal tritide.

The above minimum density ($\rho_{Stan}$) can be estimated by $\rho_{Stan} = 1/\overline{V}$, where $\overline{V}$ is the average volume of a helium atom diffusion without interaction with other helium atoms in this sphere element. This volume ($\overline{V}$) was calculated by $\overline{V} = \pi \cdot r_c^2 \cdot \overline{r_0}$, assuming a helium atom moving in a cylinder with radius ($r_c$) of 3 Å, which is the maximum length of interaction between two helium atoms. Where $\overline{r_0}$, the average diffusion length of a helium atom migrating to the centre of defects (blue area in **Fig.1**), was calculated, according to $\overline{r_0} = 3r_1/4$.

To obtain $N_{trap}$, the defect is also regarded as a sphere (blue range, radius is $r_2$), as show in **Fig.1**. In a time unit ($\Delta t$), the $N_{trap}$ is equal to the number of helium atoms ($N$) in a region nearby the surface of the defect, where helium atoms only move one step to the surface of the defect along the $\vec{r_2}$ radial direction. Obviously, the number

($dN$) of helium atoms in the rectangular box (in Fig.1) nearby the surface of the defects can be obtained via

$$dN = \Delta S \cdot \Delta L \cdot \rho(t) \tag{1}$$

Where $\Delta S$ and $\Delta L$ are bottom area and side length of the rectangular box, as show in **Fig.1**, and $\rho(t)$ is the density of helium atoms in metal tritide. The $\Delta L$ is calculated by $\Delta L = \Delta t \cdot v$, here $v$ is the velocity of a helium atom diffusion in metal tritide. Accordingly, the $N_{trap}$ is given by

$$\begin{aligned} N_{Trap} &= N \\ &= \iint c \cdot dN \cdot dS \cdot dt \\ &= \int c \cdot \rho(t) \cdot v \cdot S_{def} \cdot dt \end{aligned} \tag{2}$$

Where $S_{def}$ is surface area of the defect and $S_{def} = 4\pi r_2^2$, $c = 1/6$ is the probability of a helium atom diffusion along the $\vec{r_2}$ radial direction.

To calculate the $\rho(t)$, the rate equation of helium atoms in this sphere element (Fig. 1(b)) are established. Within a time unit ($\Delta t$), the number of added He atoms ($N_{add}$) should be equal to the number differences of between generated helium atoms ($N_{gen}$) by tritium decay and trapped helium atoms ($N_{trap}$) at the defect, as followed

$$N_{add} = N_{gen} - N_{trap} \tag{3}$$

with

$$\begin{aligned} N_{add} &= (\rho_{t+\Delta t} - \rho_t) \cdot V_{tot} \\ N_{gen} &= R(t) \cdot \Delta t \\ N_{trap} &= \frac{1}{6} \cdot v \cdot \rho(t) \cdot \Delta t \cdot S_{def} \end{aligned}$$

where $R(t)$ is the production rate of helium from tritium decay, which can be calculated by $R(t) = \alpha N_0 e^{-\alpha t}$ here $\alpha = 1.779 \times 10^{-9} \, s^{-1}$ and $N_0$ are the T-decay constant and the original amount of tritium in the small sphere element respectively. So that Eq.3 is rewritten as

$$\frac{d\rho}{dt} \cdot V_{tot} = \alpha N_0 e^{-\alpha t} - \frac{1}{6} v \rho S_{def} \tag{4}$$

Where $v = \Gamma \cdot r$ is He diffusion velocity here $\Gamma$ and $r$ are the hopping frequency and length of the migration of a helium atom in metal tritide, respectively. Then the density ($\rho(t)$) is obtained by solving the Eq. 4 in the form

$$\rho(t) = \frac{\alpha N_0}{V_{tot}(\beta - \alpha)}\left[e^{-\alpha t} - e^{-\beta t}\right] \tag{5}$$

where $\beta = \Gamma \cdot r / 2r_1$, $V_{tot} = 4\pi(r_1^3 - r_2^3)/3$. So that the number of trapped helium atoms ($N_{tra}$) at defects is given by

$$N_{tra} = \frac{1}{6}\int \Gamma \cdot r \cdot \frac{\alpha N_0}{V_{tot}(\beta - \alpha)}\left[e^{-\alpha t} - e^{-\beta t}\right] \cdot S_{def} dt \tag{6}$$

Where the volume of the sphere element ($V_{tot} = 1/\rho_{Bubble}$) is estimated by the density ($\rho_{Bubble} = 3\square 5 \times 10^{23}\, m^{-3}$) of helium bubbles in metal tritide [16]. From the Equation (5) and (6), it is shown that the calculation of the hopping frequency ($\Gamma$) is key case to apply the above model.

To obtained the frequency ($\Gamma$) of the migration of a helium atom in metal tritide, accurately, a statistical model was used in calculations based on the fact that the kinetic energy ($\varepsilon$) of a single atom in condensed matters obeys the Boltzmann distribution, $\varepsilon^{1/2} e^{-\varepsilon/k_B T}$, which has already been confirmed by a great deal of the molecular dynamics simulations[24-26]. Within a time unit, the total time for an atom to obtain a kinetic energy larger than the barrier $E_0$ ($\varepsilon > E_0$) is

$$t = \frac{1}{Z}\int_{E_0}^{\infty} \varepsilon^{1/2} e^{-\varepsilon/k_B T} d\varepsilon \tag{7}$$

Where $Z = \int_0^\infty \varepsilon^{1/2} e^{-\varepsilon/k_B T} d\varepsilon = \sqrt{\pi}(k_B T)^{3/2}/2$ is the partition function. Considering an atom located at the bottom of a potential well V(x) with kinetic energy ε (ε > E$_0$), the time taken by this atom to escape from the valley is $\delta t = \sqrt{m} \int_0^b dx / \sqrt{2(\varepsilon - V(x))}$, where *b* is the half width of the well and the average time at a certain temperature *T* can be obtained by

$$\overline{\delta t} = \frac{\int_{E_0}^\infty (\delta t) \varepsilon^{1/2} e^{-\varepsilon/k_B T} d\varepsilon}{\int_{E_0}^\infty \varepsilon^{1/2} e^{-\varepsilon/k_B T} d\varepsilon} \tag{8}$$

So the frequency (or rate) of the hopping event is

$$\Gamma = \frac{t}{\overline{\delta t}} = \frac{1}{Z} \cdot \frac{\left(\int_{E_0}^\infty \varepsilon^{1/2} e^{-\varepsilon/k_B T} d\varepsilon\right)^2}{\int_{E_0}^\infty (\delta t) \varepsilon^{1/2} e^{-\varepsilon/k_B T} d\varepsilon} \tag{9}$$

To apply this model for predicting the frequency of the hopping of a helium atom in metal tritide, we calculate the relevant potential energy curve (PEC) along the minimum energy path (MEP). Then we can use Eq. 9 to calculate the frequency (Γ) at any temperature.

To get the accurate MEP, the first principles calculations were performed to obtain the equilibrium lattice parameters and optimize the configuration, using the Vienna Ab Initio simulation Package (VASP) code at the level of Perdew-Burke-Ernzrehof parameterized generalized-gradient approximation. Projector augmented-wave pseudo potentials were used [27, 28] with a cutoff energy of 400 eV. And then the climbing image nudged elastic band (CI-NEB) method [29] embedded in the VASP code was used to investigate the migration mechanisms and energy barriers of helium diffusions. The migration barrier energy is the difference between this transition state total energy and the total energy of the nearby local minima. A 4×4×4

Monkhorst-Pack grid was used to sample the Brillouin zone. The forces on all atoms in each image of the CI-NEB chain were converged to 0.01 eV.

**Applications**

It would be ideal if this mode of aging related material degradation could be precisely predicted, and as such it is important to understand the fundamental atomistic behavior of helium in metal tritide and bubbles nucleation and growth mechanism. In order to test our theory models, take ideal stoichiometric $\beta$-TiT$_2$ as, we first performed DFT calculations to study the diffusion of a helium atom in titanium tritide, which is used commonly in nuclear industry. The $\beta$-phase of titanium tritide (TiT$_2$) has an ideal gas/metal ratio of 2.0 and the face-centered cubic (fcc) structure, where tritium occupies all of the tetrahedral interstitial sites while the octahedral interstices remain unoccupied. Since the electronic structures of hydrogen and tritium, helium and $^3$helium, are almost the same, it is reasonable to use hydrogen and helium as replacements to simulate the electronic behavior of tritium and $^3$helium in metal tritides, respectively. Figure 2 shows a conventional unit cell of $\beta$-TiT$_2$ with tetrahedral tritium (gray atoms) and fcc titanium atoms (orange atoms). The computed equilibrium lattice constant for $\beta$-phase TiT$_2$ is 4.424 Å, which is consistent with the experimental value of 4.440 Å [30]. A 2×2×1 supercell of 48 atoms was used to represent the ground state $\beta$-TiT$_2$ crystal, where one tritium atom was replaced by a helium atom.

In material containing helium from tritium decay, there is initially a tetrahedral vacancy (V$_{tet}$) associated with every helium atom produced. The newly created helium atom could occupy the tetrahedral site vacated by a decayed tritium, or it could prefer one of the nearest-neighboring octahedral sites (V$_{oct}$). Our calculations predict that helium will occupy the tetrahedral site ($V_{tet}$) rather than the nearest-neighboring octahedral site ($V_{oct}$) and is not stable in an octahedral site neighboring an empty tetrahedral site, similar to that calculated in ErH$_2$ [20]. Then, in ideal stoichiometric $\beta$-TiT$_2$, the helium atom occupied tetrahedral site may have two migration paths as shown in Fig. 2. One is that migrating via exchanging site with the nearest tetrahedral tritium atoms (Fig. 2(a)). Another is that moving to a next-nearest-neighbor octahedral site with nearest-neighbor filled tetrahedral site (Fig.2(b)) and then migrating between the octahedral sites (Fig. 2(c)). The CI-NEB results for helium migration in $\beta$-TiT$_2$ are shown in Fig.3 and illustrated in Fig.2. Energies, obtained with the NEB method, of the migration path and barrier for displacement of the nearest tetrahedral tritium by helium are shown in Fig. 3(a). The migration by this mechanism has a lower rate-limiting barrier (0.81eV) than that (1.85 eV) in the conceptually straightforward mechanism of tetrahedral to next-nearest-neighbor octahedral transitions shown in Fig. 3(b), and will be the more dominant mechanism for helium diffusion in $\beta$-TiT$_2$.

Then, the frequency of the hopping of a helium atom in this case was obtained, using the statistical model of a single atom (Eq.9). The calculated frequency $\Gamma$ is 0.96 s$^{-1}$ at room temperature, suggesting the helium atoms easily move in $\beta$-TiT$_2$.

Finally, the helium bubble formation was described by our model (Eq.4 and Eq.5) in titanium tritide. The density of helium atoms ($\rho$), the number of rapped helium atoms ($N_{tra}$) and the diameter of helium bubble as a function of aging time at room temperature, in this sphere model, were shown in Fig. 4. The max value of density ($\rho_{max}$) is $2.96 \times 10^{24}$ m$^{-3}$ at room temperature, which is two orders of magnitude smaller than $\rho_{Stan}$ (~$6.05 \times 10^{26}$ m$^{-3}$), suggesting that the supposition of non-interaction between helium atoms is reasonable during them diffusion. In Fig. 4(a), it is noted that the $N_{tra}$ increase linearly with aging time increasing at 300K, meaning the diameters of helium bubble centered on defects in titanium tritide increase with $t^{1/3}$(Fig. 4(b)), here $t$ is aging time, on the basis of the equation as follows

$$V_{Bubble} = 4\pi r^3_{Bubble}/3 = \rho' N_{tra}$$

where $\rho'$ is the He density in such bubbles that is approximately constant (~$1.3 \times 10^{29}$ m$^{-3}$) corresponding to pressures between 5 and 10 GPa [9] and $r_{Bubble}$ is radius of helium bubbles. This result agrees quite well with the experimental phenomenon [9, 13], suggesting our model can provide a reasonable physical picture of helium bubbles nucleation and growth in metal tritide. Thus we suggest the interaction between helium atoms, forming He clusters that act as nucleating center for bubble formation, should been neglected on the study of helium microcosmic diffusion mechanism in metal tritides at room temperature. It is worth to note that the calculated diameter of bubbles shown in Fig.4 (b) by the above model was bigger than that in experiments [13] at the same aging time, for example it is ~4 nm using our model and only ~2.5 nm in experiment after 180 days aging, because this model neglects the presence of impurities and micron scale grain boundaries or variations in stoichiometry in actual titanium trtides. Nevertheless, our model will serve as a basis for interpretation of the helium bubble nucleation and serve as a look into the atomistic processes in metal tritides.

It is well known that the temperatures and energy barrier of a He atom migration on the MEP may be two critical factors to prolong the service lifetime of metal tritide by restraining helium atoms diffusion and bubbles growth. Unfortunately, experimental studies on the influence of these factors are extremely time consuming and difficult, while there is a void in theoretical research on influence of these factors. It's worth noting that our model can have quantitative insight into the impact of these factors. Firstly, to discuss the influence of temperatures on the helium bubble growth in titanium tritide, the variations of helium bubbles diameter as aging time increases in temperature rang 200 to 325K were shown in Fig. 5. The results show that growth of helium bubbles was obviously restrained below 225K. For instance, the bubble diameter (~1.5nm) at 225K is smaller than that (~5.0nm) at 300K as the aging time is 300 days.

Secondly, for the influence of energy barrier on the helium bubbles growth in titanium tritide, the variations of bubble diameter with aging time increasing were calculated in energy barrier rang 0.7 to 1.2 eV, plotted in the Fig. 6, which shows that the growth of helium bubbles is obviously restrained when the energy barriers exceed 1.1 eV, only more than 0.3 eV to the one calculated (0.81 eV). Clearly, the bubble growth is strongly dependent on the energy barriers of a He atom migration on an MEP, suggesting that the way of increasing energy barriers should be effective against bubble growth.

Interestingly, the helium accelerated release from titanium tritides, leading to the deteriorated mechanical properties of titanium tritides in general case, as the helium and titanium atom ratio is up to about 0.3 (~3years) at room temperature [31]. The calculated helium bubble diameter in titanium tritides was 7.5 nm using our model at 300 K (with an energy barrier of 0.81eV) after 3 years. however, the calculations show that it needs ~12 years to obtain a 7.5nm-diameter bubble at 225K (with the same energy barrier of 0.81 eV) or ~16 years with an energy barrier of 1.1eV (at the same temperature of 300K). It is shown that the temperature of 225 K used or the way of increasing diffusion barrier upto 1.1 eV is adequate to prolong life of titanium tritides, contrast with the half-life of tritium decays is 12.32 years.

**Conclusions**

The model of helium bubbles nucleation and growth in metal tritide was expected to uncover fundamental physical mechanisms and provide ways to prolong service life of metal tritide, based on a simple model, suggesting helium bubbles nucleate and growth at inherent defects in metal tritide, the bulk metal tritide are composed of the small sphere element centered on inherent defects, helium uniform distribute, the density of helium bubbles is consistent with that of inherent defects and non-interaction between helium atoms in the process of diffusion in metal tritide. In order to test our model, take ideal stoichiometric $\beta$-TiT$_2$ as, first, the migration mechanisms of He were investigated by using ab initio calculation based on density functional theory. The diffusion of helium can proceed by transition to a tetrahedral site and continuing to migrate via exchanging site with the nearest tetrahedral tritium atom with an energy barrier of 0.81 eV. Then, the diameter of the helium bubble was calculated using our model, which increases linearly as $t^{1/3}$ for titanium tritide, agreed quite well with the experimental phenomenon. The results show that our model can provide a reasonable physical mechanism of helium bubbles nucleation and growth and suggest that interaction between helium atoms, forming He clusters, should been neglected on the study of helium microcosmic diffusion mechanism in metal tritides at room temperature. Two critical factors (the temperature and diffusion barrier of a helium atom on the MEP) to prolong the service life of metal tritides by restraining helium atoms diffusion and bubbles growth were discussed quantificationally using our model. The way of reducing temperature to 225 K or increasing diffusion barrier upto 1.1 eV can effectively restrain bubbles growth and prolong lifetime of titanium tritides, which provide a useful reference to the experiment and have important implications in the development of new tritium-storage materials and plasma facing materials in nuclear industry.

**Reference**
[1] R. S. Barnes. Nature, 206, 1307~1310 (1965)
[2] W.J.Camp, J. Vac. Sci.Technol. 14,514 (1977)
[3] R. Lässer, Tritium and Helium-3 in Metals, Springer-Verlag, Berlin, 1989
[4] R. G. Spulak, Jr., J.Less-Common Met. 132, L17 (1987)
[5] G. J. Thomas, Radiat. Eff. 78, 37 (1983)
[6] Y.Katoh, M. Ando, and A. Kohyama, J.Nucl.Mater. 323 251-262. (2003)
[7] J.A.Knapp and J. F. Browning, J.Nucl.Mater.350, 147. (2006)
[8] J.A.Knapp, D. M. Follstaedt, and S. M. Myers, J. Appl. Phys.103, 013518 (2008)
[9] T. Schober, H.Trinkaus, and R. Lässer, J.Nucl.Mater.141, 453-457(1986)
[10] T. Schober, R. Lässer, W.Jäger, and G. J. Thomas, J. Nucl. Mater., 122, 571-575 (1984)
[11] T. Schober and R. Lässer, J.Nucl.Mater.120, 137(1984)
[12] T. Schober and H.Trinkaus, Philos. Mag. A, 65,1235-1247(1992)
[13] T. Schober and K. Farrell, J.Nucl.Mater. 168, 171-177(1989)

[14] G. C. Abell, L. K. Matson, and R. H. Steinmeyer, Phys. Rew. B, 41,1220-1223(1990)

[15] J. A. Knapp, J. F. Browning, and G. M. Bond, J. Appl. Phys., 105, 053501(2009)

[16] G. M. Bond, J. F. Browning, and C. S. Snow, J. Appl. Phys.,107, 083514(2010)

[17] H. T. Weaver, W. J. Camp, Phys. Rev. B, 12,3054 (1975)

[18] W.D. Wilson, C. L. Bisson, and M. I. Baskes, Phys. Rev. B, 24,5616-5624 (1981)

[19] R. Vassen, H. Trinkaus, and P. Jung, Phys. Rev. B, 44,4206 (1991)

[20] R. R. Wixom, J. F. Browning, C. S. Snown, P. A. Schultz, and D. R. Jennison, J. Appl. Phys., 103,123708 (2008)

[21] C. S. Becquart and C. Domain, Phys. Rev. Lett., 97,196402(2006)

[22] T. Seletskaia, Y. Qsetsky, R. E. Stoller, and G. M. Stocks, Phys. Rev. B, 78, 134103(2008)

[23] P. B. Zhang, J. J. Zhao, Y. Qin, and B. Wen, J.Nucl.Mater.419, 1-8(2011)

[24] Z. Lin, W. Yu, Y. Wang, and X. Ning, EPL 94, 40002 (2011)

[25] C. Ming, Z.-Z. Lin, R.-G. Cao,W.-F. Yu, and X.-J. Ning, Carbon,50, 2651 (2012).

[26] Wei-Feng Yu, Zheng-Zhe Lin, and Xi-Jing Ning, Chin. Phys. B, 22, 116802 (2013)

[27] G. Kresse, D. Joubert, Phys. Rev. B, 59, 1758-1775(1999)

[28] P. E. Blochl, Phys. Rev. B, 50,17953-17979(1994)

[29] G. Henkelman, B. Uberuaga, and H. Jonsson, J. Chem. Phys.113,9901-9904(2000)

[30] J.F. Fernandez, F. Cuevas, C. Sanchez, J. Alloy. Compd. 205,303-309(1994)

[31] R.P. Gupta, M. Gupta, Phys. Rev. B, 66,014105(2002)

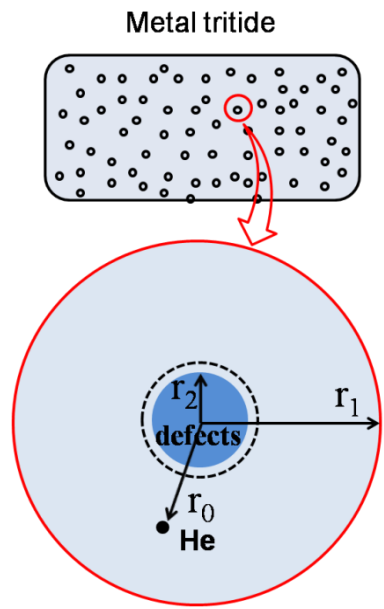

Fig. 1. (a) the helium bubble in metal tritide and (b) the small sphere element centered on defects

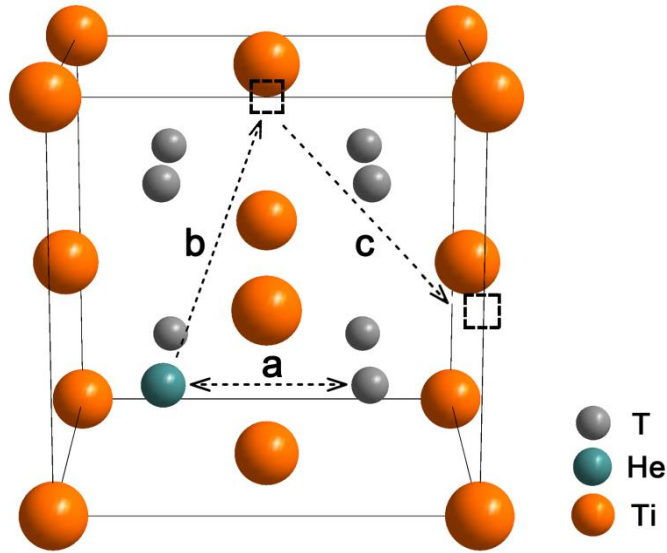

Fig. 2. A unit cell of $\beta$-TiT$_2$ and illustration of helium migration mechanisms. The larger orange and smaller gray atoms represent the titanium fcc lattice and tritium occupying all of the tetrahedral site, respectively. The teal atoms represent helium from tritium decay. The octahedral sites, in the middle of each cube edge and in the cube center, are unoccupied. (a) Helium migration by exchanging site with the nearest direct tetrahedral tritium atoms. (b) Migration of helium moving to a next-nearest-neighbor octahedral site, and then (c) migrating between the octahedral sites.

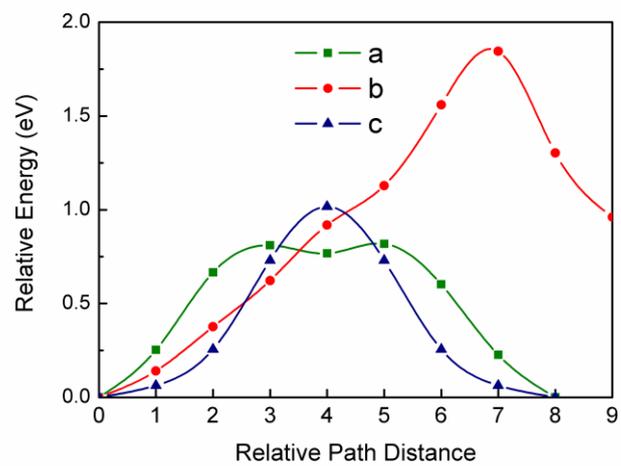

Fig. 3. CI-NEB data for helium migration in TiT$_2$ corresponding to the migration paths in fig.2.

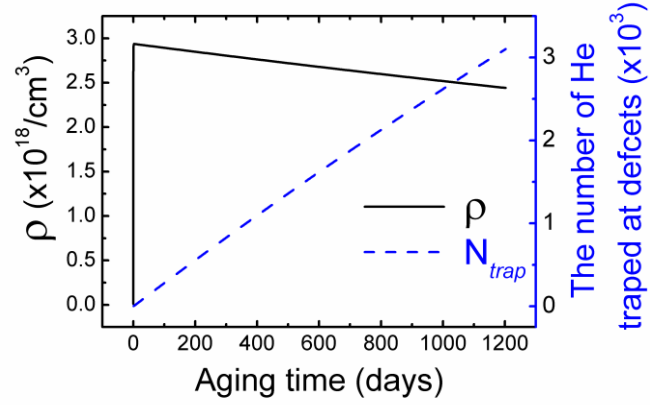

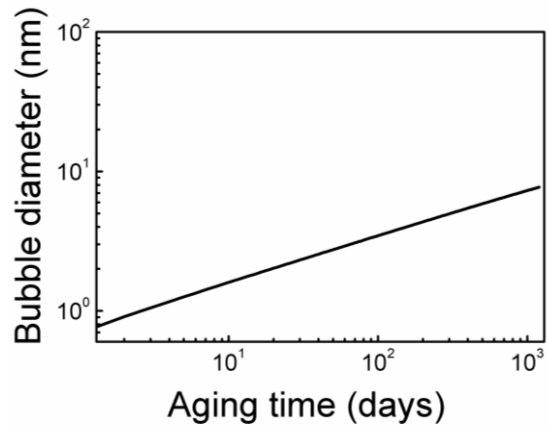

Fig. 4. (a) The density of helium atoms ($\rho$) and the number of rapped helium atoms ($N_{tra}$) at defects as a function of aging time at room temperature, in this sphere model. (b) The diameter of helium bubble centered on defects as a function of aging time.

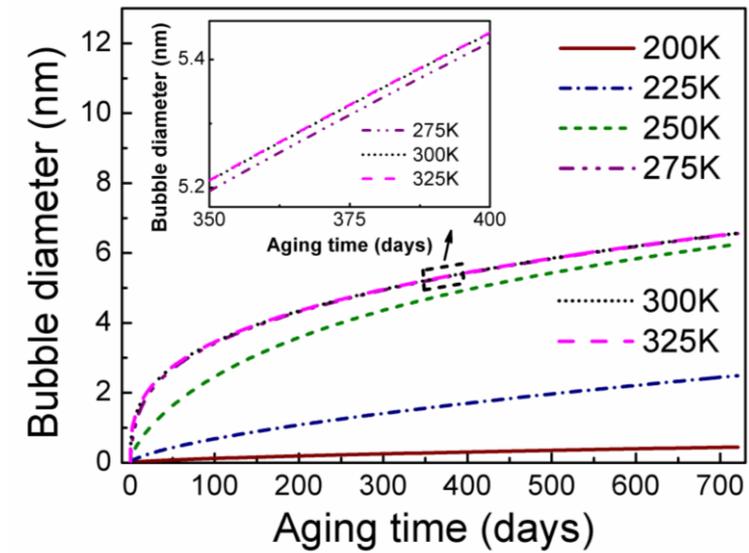

Fig.5. The diameter of helium bubble as a function of aging time in temperature rang 200 to 325K.

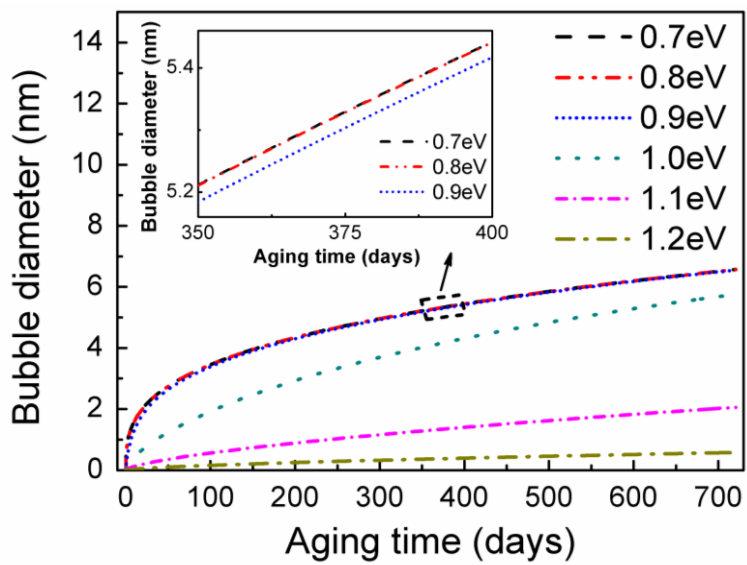

Fig. 6. The diameter of helium bubble as a function of aging time in energy barrier rang 0.7 to 1.2 eV